\documentclass[10pt,final,doubleside]{IEEEtran}
	\usepackage{pifont,bm,multicol,amsfonts,amsmath,color,amssymb,graphicx, epsfig,cite,psfrag,subfigure,algorithm,enumerate,stfloats,algorithm,algorithmic,epstopdf,balance,multirow,multicol,makecell}

\begin{document}

\title{Deep Learning Based Antenna-time Domain Channel Extrapolation for Hybrid mmWave Massive MIMO }

\author{Shunbo Zhang, Shun Zhang, \emph{Senior Member, IEEE}, Jianpeng Ma, \emph{Member, IEEE}, Tian Liu, Octavia A. Dobre, \emph{Fellow, IEEE}
    \thanks{S. Zhang, S. Zhang and J. Ma are with the State Key Laboratory of Integrated Services Networks, Xidian University, Xi'an 710071, P. R. China (e-mail: sbzhang$\_$19@stu.xidian.edu.cn, zhangshunsdu@xidian.edu.cn, jpmaxdu@gmail.com).

	T. Liu is with the Science and Technology on Electronic Information Control Laboratory, Chengdu 610036, P. R. China (e-mail: xidianlt@163.com).

    O. A. Dobre is with Faculty of Engineering and Applied Science, Memorial University, St. John's NL AIC-5S7, Canada (e-mail: odobre@mun.ca).}	
}

\maketitle

\vspace{-10mm}
\begin{abstract}
	In a time-varying massive multiple-input multiple-output (MIMO) system, the acquisition of the downlink channel state information at the base station (BS) is a very challenging task due to the prohibitively high overheads associated with downlink training and uplink feedback.
	In this paper, we consider the hybrid precoding structure at BS and examine the antenna-time domain channel extrapolation.
	We design a latent ordinary differential equation (ODE)-based network under the variational auto-encoder (VAE) framework to learn the mapping function from the partial uplink channels to the full downlink ones at the BS side.
	Specifically, the gated recurrent unit is adopted for the encoder and the fully-connected neural network is used for the decoder. The end-to-end learning is utilized to optimize the network parameters.
	Simulation results show that the designed network can efficiently infer the full downlink channels from the partial uplink ones, which can significantly reduce the channel training overhead.
\end{abstract}

\thispagestyle{empty}
\vspace{-1mm}



\section{Introduction}
\label{Introduction}

Due to its significant improvement in spectral and energy efficiencies, massive multiple-input multiple-output (MIMO) has become a key technology in the future wireless communication network \cite{massive_mimo1}.
To fully exploit the advantages of the massive MIMO systems, accurate downlink channel state information (CSI) should be obtained at the base station (BS) for precoding.
However, the downlink CSI acquisition is a very challenging task in the massive MIMO systems, which requires overwhelming channel training overhead, especially in the time-varying channel scenario \cite{noh_time_varying,ma_time_varying}.

There are many works focus on the time-varying massive MIMO channel acquisition \cite{qin_time_varying, han_time_varying, xia_time_varying}.
In \cite{qin_time_varying}, Qin \emph{et al.}
developed a quasi-block simultaneous orthogonal matching pursuit algorithm to recover the time-varying massive MIMO channels.
In \cite{han_time_varying}, by leveraging the reciprocity between the uplink and downlink channels, Han \emph{et al.} adopted the extended Newtonized orthogonal matching pursuit algorithm to extract the delay and angular parameters from the received uplink pilots, and designed a module to trace the downlink gains through channel training and feedback.
In \cite{xia_time_varying}, Xia \emph{et al.} analyzed the spatial-temporal sparse structure of the time-varying massive MIMO channel and formulated a structured variational Bayesian inference framework to conduct the channel estimation.
All these works \cite{qin_time_varying, han_time_varying, xia_time_varying} are closely dependent on hypothetical mathematical models and rely on the iterative algorithms.
However, in the practical communication scenario, the radio scattering conditions can be very complicated, which may cause serious mismatch between the hypothetical model and the actual one.

Recently, deep learning (DL) has been introduced to the massive MIMO systems due to its excellent performance and low complexity.
In \cite{wang_dl_channel}, Wang \emph{et al.} proposed a real-time CSI feedback framework by extending the DL-based CSI network with long short-term memory and achieved a remarkable recovery quality of the time-varying massive MIMO channel.
In \cite{yang_dl_channel}, Yang \emph{et al.} resorted to the graph neural network (NN) and designed a novel massive MIMO channel tracking framework, which achieved better performance than that with feedforward NN under the high mobility scenario.
Yang \emph{et al.} designed a deep transfer learning-based downlink channel prediction network for frequency division duplexing massive MIMO systems, and a meta-learning algorithm was proposed to train the network and then adapt to a new environment with a small number of labeled data \cite{yangyu_dl_channel}.

In this paper, we consider the DL-based antenna-time domain channel extrapolation for the time-varying massive MIMO systems.
Specifically, the hybrid precoding structure is adopted to reduce the number of the active antennas at the BS.
We first investigate the existence of the mapping function from the partial uplink channels to the full downlink ones.
After that, we resort to the variational auto-encoder (VAE) framework and the latent ordinary differential equation (ODE) model to design the channel extrapolation network as the implementation of the mapping function, where the gated recurrent unit (GRU) is used for the encoder and the fully-connected NN (FNN) is adopted for the decoder.
Lastly, we utilize the end-to-end learning to optimize the parameters of the designed network.

\emph{Notations:}
We use lowercase (uppercase) boldface to denote vector (matrix).
$(\cdot)^T$, $(\cdot)^H$, $\|\cdot\|_F$ and $\mathbb{E}\{\cdot\}$ represent the transpose, Hermitian, Frobenius norm and expectation, respectively.
$\odot$ is the Hadamard product operator, $\circ$ is the composite mapping operator, and $|\mathcal{A}|$ is the number of elements in set $\mathcal{A}$.
$v\sim \mathcal{N}(\mu,\sigma^2)$ means that scalar $v$ follows the Gaussian distribution with mean $\mu$ and variance $\sigma^2$.
$\mathbf{v}\sim \mathcal{CN}(\mu,\sigma^2\mathbf{I}_N)$ means that vector $\mathbf{v}$ follows the complex Gaussian distribution with mean $\mu$ and variance $\sigma^2$.
$\mathbf{I}_N$ represents a $N\times N$ identity matrix.
$[\mathbf{a};\mathbf{b}]$ is a vector that connects two vectors $\mathbf{a}$ and $\mathbf{b}$.

\section{System Model}
\label{System_Model}

\subsection{Time-varying Massive MIMO Channel Model}
\label{Channel_model}

\begin{figure}
	\centering
	\includegraphics[width=80mm]{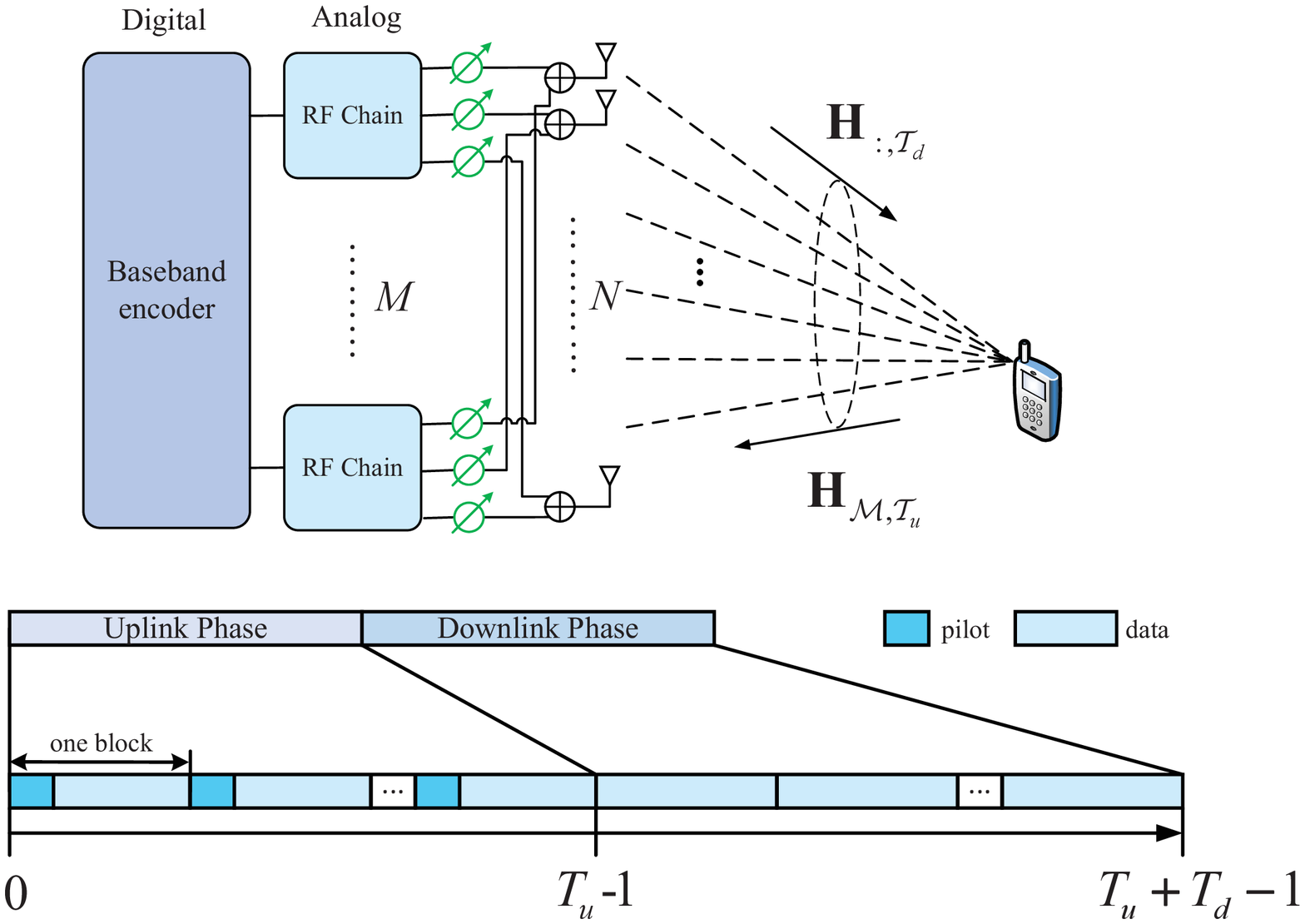}
	\caption{Hybrid structure and the block diagram of each communication process between the BS and the user.}
	\label{hybrid_frame}
\end{figure}

Consider a massive MIMO system, which contains one BS and one user.
The BS is equipped with $N\gg 1$ antennas in the form of uniform linear array (ULA), and the user is equipped with single antenna.
As depicted in Fig. \ref{hybrid_frame}, to decrease the hardware cost, the BS adopts the hybrid precoding structure \cite{zhao_twc} and contains $M$ radio frequency (RF) chains, where $M\ll N$.
Define the spatial compression ratio as $r=\frac{M}{N}$.
Without loss of generality, we consider the time-division duplex (TDD) mode, where the channel reciprocity exists.
As shown in Fig. \ref{hybrid_frame}, each communication process between the BS and the user is divided into two phases, i.e., the uplink and downlink ones.
The former contains $T_u$ time blocks, while the latter consists of $T_d$ time blocks.
The antenna-time domain communication resource can be written as the set $\mathcal C=\{(i,n)|i\in\mathcal{N}=\{0,1,2,\ldots,N-1\},n\in\mathcal T=\{0,1,2,\ldots,T_u+T_d-1\}\}$, where $i$ is the antenna index, and $n$ denotes the time block index. 
Notice that $n\in\mathcal T_{u}=\{0,1,\ldots,T_u-1\}$ and $n\in\mathcal T_d=\{T_u,T_u+1,\ldots,T_u+T_d-1\}$ correspond to the uplink phase and the downlink one, respectively.

Assume that the channel is quasi-static during a time block of $L_c$ channel uses and changes from block to block.
Define the $N\times (T_u+T_d)$ matrix $\mathbf{H}$ as the antenna-time domain channel between the BS and the user.
During the $n$-th time block of the communication process, the $N\times 1$ time-varying channel vector between the user and the BS is expressed as
\begin{align}
\label{eq:channel_model}
	\mathbf{H}_{:,n}=\sum_{p=1}^{N_p}\alpha_{p} e^{\jmath2\pi(\nu_p nL_cT_s-f_c\tau_p)}\mathbf{a}({\theta}_p(n)),
\end{align}
where $N_p$ is the number of channel scattering paths, $T_s$ is the sampling period, $f_c$ is the carrier frequency, $\alpha_p$, $\nu_p$ and $\tau_p$ denote the complex channel gain, the Doppler shift and the delay of the $p$-th path, respectively.
Moreover, the spatial steering vector $\mathbf{a}(\theta_p(n))$ is defined as
\begin{align}
\label{eq:steering_vector}
	\mathbf{a}(\theta_p(n))=\left[1,e^{\jmath 2\pi\frac{d}{\lambda}\sin\theta_p(n)},\cdots,e^{\jmath 2\pi\frac{d}{\lambda}(N-1)\sin\theta_p(n)}\right]^T,
\end{align}
where $d$ is the antenna spacing, $\lambda$ denotes the carrier wavelength and $\theta_p(n)$ represents the angle of arrival/departure (AoA/AoD) for the $p$-th path of the channel at time block $n$.
As $\theta_p(n)$ remains constant during a long time interval, the time index $n$ of the AoA/AoD can be omitted.

\section{DL-based Antenna-time Domain Channel Extrapolation}
\label{Extrapolation}

\subsection{Problem Description}
\label{Problem}

In order to implement effective uplink data detection and downlink precoding at the BS, $\mathbf{H}$ should be obtained as accurate as possible, which is a challenging task.
In the analog-digital hybrid structure, only the uplink CSI at $M$ instead of $N$ antennas can be obtained at the same time.
Assume that these $M$ antennas are separately selected to be connected to the RF chains at the BS.
The indexes of the selected antennas keep unchanged within $\mathcal{T}_u$ and are collected into the set  $\mathcal{M}$.
As depicted in Fig. \ref{hybrid_frame}, during the uplink phase, the user sends the pilot signal $x_n$ at time index $n\in\mathcal{T}_u$, where $\mathbb{E}\{|x_n|^2\}=P$ and $P$ denotes the transmit power.
The corresponding $M\times 1$ received signal at the BS is expressed as 
\begin{align}
\label{eq:received}
	\mathbf{y}=\mathbf{H}_{\mathcal{M},n}x_n+\mathbf{n}_n,
\end{align}
where $\mathbf{H}_{\mathcal{M},n}$ is the uplink partial channel vector at time index $n$ and $\mathbf{n}_n\sim\mathcal{CN}(0,\sigma^2\mathbf{I}_M)$ denotes the additive white Gaussian noise vector.
The signal-to-noise (SNR) is defined as $\text{SNR}=P/\sigma^2$.
By leveraging the least square estimation for each time index $n$, the $M\times T_u$ uplink partial channel matrix $\mathbf{H}_{\mathcal{M},\mathcal{T}_u}$ can be explicitly obtained at the BS with sufficient SNR.

For the given communication environment and the fixed array structure, the antenna domain mapping function from the partial uplink channel $\mathbf{H}_{\mathcal{M},\mathcal{T}_u}$ to the $N\times T_u$ full uplink channel $\mathbf{H}_{:,\mathcal{T}_u}$ can be expressed as \cite{channel_mapping,zhang_tcom}
\begin{align}
\label{eq:ant_map}
	\bm{\Phi}_{\mathcal{M}\rightarrow \mathcal{N}}^a:\{\mathbf{H}_{\mathcal{M},\mathcal{T}_u}\}\rightarrow  \{\mathbf{H}_{:,\mathcal{T}_u}\}.
\end{align}
With the above mapping \eqref{eq:ant_map}, we can perform the antenna domain extrapolation with respect to $\mathbf{H}_{\mathcal{M},\mathcal{T}_u}$ and obtain the accurate estimation of $\mathbf{H}_{:,\mathcal{T}_u}$.

In the TDD mode, there exists reciprocity between uplink and downlink channels, which is usually utilized to decrease
the pilot overhead for the downlink massive MIMO channels estimation \cite{xie_twc}.
However, in the time-varying channel scenario, the uplink and downlink channels are not exactly the same.
Fortunately, the movement speed of the user  and the scattering scenarios remains constant within one time interval, and it is possible to infer the $N\times T_d$ full downlink channel $\mathbf{H}_{:,\mathcal{T}_d}$ from the full uplink one $\mathbf{H}_{:,\mathcal{T}_u}$, which is due to the existence of the time domain mapping function
\begin{align}
\label{eq:time_map}
	\bm{\Phi}_{\mathcal{T}_u\rightarrow \mathcal{T}_d}^t:\{\mathbf H_{:,\mathcal T_u}\}\rightarrow  \{\mathbf H_{:,\mathcal T_d}\}.
\end{align}

With \eqref{eq:ant_map} and \eqref{eq:time_map}, we can obtain the mapping function from $\mathbf{H}_{\mathcal{M},\mathcal{T}_u}$ to $\mathbf{H}_{:,\mathcal{T}_d}$ as
\begin{align}
\label{eq:extra_map}
	\bm{\Psi}_{\mathcal{M},\mathcal{T}_u\rightarrow \mathcal{N},\mathcal{T}_d}=\bm{\Phi}_{\mathcal{M}\rightarrow \mathcal{N}}^a\circ \bm{\Phi}_{\mathcal{T}_u\rightarrow \mathcal{T}_d}^t:\{\mathbf{H}_{\mathcal{M},\mathcal{T}_u}\}\rightarrow \{\mathbf{H}_{:,\mathcal{T}_d}\}.
\end{align}
In this paper, we design a DL-based antenna-time domain channel extrapolation algorithm to complete the above mapping function \eqref{eq:extra_map}.

\subsection{Latent ODE-based Channel Extrapolation Network}
\label{Latent}

\begin{figure*}
	\centering
	\includegraphics[width=150mm]{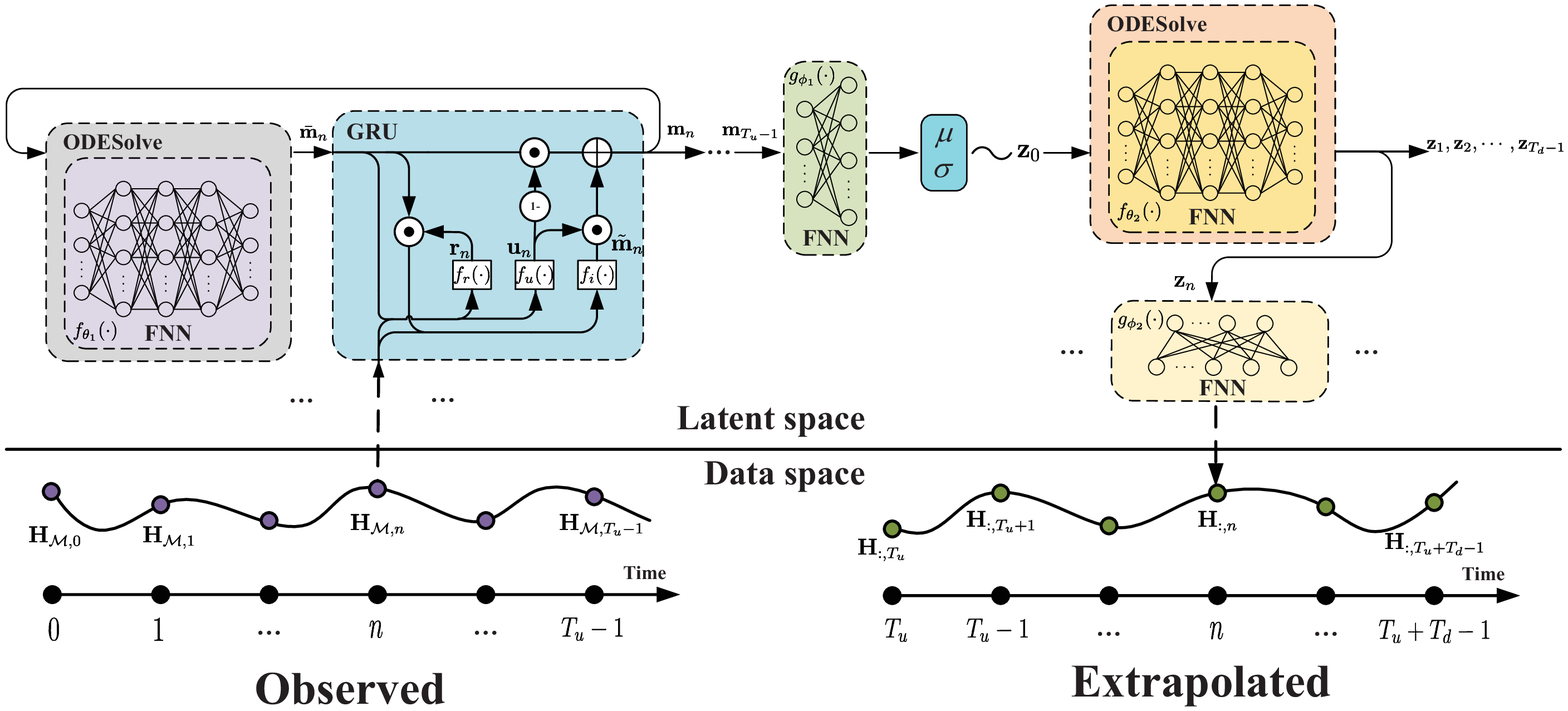}
	\caption{The structure of the designed LODE-CENet.}
	\label{latent}
\end{figure*}

To efficiently infer the downlink antenna-time domain channel, we design a \textbf{L}atent \textbf{ODE}-based \textbf{C}hannel \textbf{E}xtrapolation \textbf{Net}work (LODE-CENet) under the VAE framework, as shown in Fig. \ref{latent}.
The entire network is divided into two spaces, i.e., the data space and the latent one.
The data space contains the input and output of the LODE-CENet, where the input is the observed uplink channel $\mathbf{H}_{\mathcal{M},\mathcal{T}_u}$, and the output is the extrapolated downlink channel $\mathbf{H}_{:,\mathcal{T}_d}$.
The latent space consists of the learnable parameters and the intermediate variables of the LODE-CENet.

In order to extract the spatial and temporal characteristics of the observed channel data, we rearrange $\mathbf{H}_{\mathcal{M},\mathcal{T}_u}$ into sequences at different observation time-points as $\{\mathbf{H}_{\mathcal{M},n}\}_{n=1}^{T_u-1}$, and feed the sequences into the encoder of the LODE-CENet.
Since the encoder can be viewed as a \emph{recognition network} constructed by the ODE and the gated recurrent NN (RNN) with a GRU, we call it the ODE-RNN-based encoder.

Define $\mathbf{m}(t)\in\mathbb{R}^{L\times 1}$ as the continuous-time hidden state and $\mathbf{m}_n=\mathbf{m}(t_n)$ as the hidden state at the $n$-th observation time-point.
Unlike conventional RNN which uses the fixed hidden state between two adjacent observation time-points, we adopt an ODE to describe the evolution of the hidden state.
Specifically, the derivative of the hidden state is determined by a FNN, which is written as~\cite{latent_ODE}
\begin{align}
\label{eq:derivative}
	\frac{d\mathbf{m}(t)}{dt}=f_{\bm{\theta}_1}(\mathbf{m}(t)),
\end{align}
where $f_{\bm{\theta}_1}(\cdot)$ represents a FNN with parameters $\bm{\theta}_1$ and specifies the time-invariant dynamics of the hidden state.
By leveraging a numerical ODE solver, e.g., the Euler method and the Runge-Kutta method, the hidden state can be obtained at any expected time.

With the observed channel data, the GRU further updates the hidden state. 
Thus, the entire updating process of the hidden state from the $(n-1)$-th observation time-point to the $n$-th one can be expressed as
\begin{align}
\label{eq:ode_solver1}
	\bar{\mathbf{m}}_n&=\text{ODESolve}(f_{\bm{\theta}_1},\mathbf{m}_{n-1},(t_{n-1},t_n)),\\
\label{eq:gru}
	\mathbf{m}_n&=\text{GRU}_{\bm{\psi}}(\bar{\mathbf{m}}_{n},\mathbf{H}_{\mathcal{M},n}),
\end{align}
where $\text{ODESolve}(\cdot)$ represents the numerical ODE solver, $\bar{\mathbf{m}}_n\in\mathbb{R}^{L\times 1}$ is the solution of the ODE \eqref{eq:derivative} given the initial value $\mathbf{m}_{n-1}$ and the time step $(t_{n-1},t_n)$, and $\text{GRU}_{\bm{\psi}}(\cdot)$ denotes the hidden state update function of the GRU with parameters $\bm{\psi}$.
Unlike standard RNN, the GRU utilizes two gates, i.e., the update gate and the reset gate, to control the update of the hidden state.
The update process of the hidden state in the GRU at the $n$-th observation time-point is presented in \textbf{Algorithm 1}, where $f_u(\cdot)$, $f_r(\cdot)$, $f_i(\cdot)$ respectively represent three FNNs.

\linespread{1}
\begin{algorithm}[t]
\caption{Hidden state update process in the GRU at the $n$-th observation time-point.}
\begin{algorithmic}[1]
	\STATE \textbf{Input:} ODE solution $\bar{\mathbf{m}}_{n}$, observation $\mathbf{H}_{\mathcal{M},n}$
	\STATE $\mathbf{u}_n=f_u([\bar{\mathbf{m}}_{n};\mathbf{H}_{\mathcal{M},n}])$, where $\mathbf{u}_n\in\mathbb{R}^{L\times 1}$ represents update coefficient
	\STATE $\mathbf{r}_n=f_r([\bar{\mathbf{m}}_{n};\mathbf{H}_{\mathcal{M},n}])$, where $\mathbf{r}_n\in\mathbb{R}^{L\times 1}$ represents reset coefficient
	\STATE $\tilde{\mathbf{m}}_n=f_i([\bar{\mathbf{m}}_{n}\odot\mathbf{r}_n;\mathbf{H}_{\mathcal{M},n}])$, where $\tilde{\mathbf{m}}_n\in\mathbb{R}^{L\times 1}$ represents intermediate hidden state
	\STATE $\mathbf{m}_n=(1-\mathbf{u}_n)\odot\tilde{\mathbf{m}}_{n}+\mathbf{u}_n\odot\bar{\mathbf{m}}_{n}$
\STATE \textbf{Return:} the new hidden state $\mathbf{m}_n$
\end{algorithmic}
\end{algorithm}

The decoder of the LODE-CENet is a \emph{generator network} based on another ODE and the FNN.
Unlike the conventional auto-encoder, which converts the input into a codeword, the ODE-RNN-based encoder estimates the approximate posterior $q(\mathbf{z}_0|\{\mathbf{H}_{\mathcal{M},n},t_n\}_{n=1}^{T_u-1})$ to enhance the generation ability of the ODE-FNN-based decoder under the VAE framework, where $\mathbf{z}_0\in\mathbb{R}^{L\times 1}$ represents the initial latent state of the ODE for the decoder and also determines the entire trajectory of the extrapolation.
We transform the final hidden state $\mathbf{m}_{T_u-1}$ to the mean vector $\bm{\mu}_{\mathbf{z}_0}\in\mathbb{R}^{L\times 1}$ and the standard deviation vector $\bm{\sigma}_{\mathbf{z}_0}\in\mathbb{R}^{L\times 1}$ through a FNN $g_{\bm{\phi}_1}(\cdot)$ with parameters $\bm{\phi}_1$, which can be written as
\begin{align}
\label{eq:transform_NN}
	[\bm{\mu}_{\mathbf{z}_0}^T,\bm{\sigma}_{\mathbf{z}_0}^T]^T=g_{\bm{\phi}_1}(\mathbf{m}_{T_u-1}).
\end{align}
Then, the approximate posterior is equal to a Gaussian distribution as
\begin{align}
\label{eq:gaussian}
	q(\mathbf{z}_0(l)|\{\mathbf{H}_{\mathcal{M},n},t_n\}_{n=1}^{T_u-1})&=\mathcal{N}(\bm{\mu}_{\mathbf{z}_0}[l],\bm{\sigma}_{\mathbf{z}_0}^2[l]),\notag\\ \text{for}\ l&=0,1,\cdots,L-1.
\end{align}
By leveraging the \emph{reparameterization trick}, $\mathbf{z}_0$ can be obtained from \eqref{eq:gaussian} as
\begin{align}
\label{eq:initial_latent}
	\mathbf{z}_0=\bm{\mu}_{\mathbf{z}_0}+\bm{\sigma}_{\mathbf{z}_0}\odot\bm{\epsilon},
\end{align}
where $\bm{\epsilon}\in\mathbb{R}^{L\times 1}$ and each element in $\bm{\epsilon}$ follows the Gaussian distribution $\mathcal{N}(0,1)$.

Similar to \eqref{eq:derivative}, we also adopt a FNN $f_{\bm{\theta}_2}(\cdot)$ with parameters $\bm{\theta}_2$ to determine the derivative of the latent state.
Define $\mathbf{z}_n$ as the latent state at the $n$-th time-point.
Then, with $\mathbf{z}_0$ and the time steps $(t_{T_u},t_{T_u+1},\cdots,t_{T_u+T_d-1})$, we can extrapolate the latent states as
\begin{align}
\label{eq:extrapolate}
	&\mathbf{z}_1,\mathbf{z}_2,\cdots,\mathbf{z}_{T_d-1}\notag\\
	&=\text{ODESolve}(f_{\bm{\theta}_2},\mathbf{z}_0,(t_{T_u},t_{T_u+1},\cdots,t_{T_u+T_d-1})).
\end{align}
By sending these latent states into a FNN $g_{\bm{\phi}_2}(\cdot)$ with parameters $\bm{\phi}_2$, the extrapolated downlink channels is obtained as
\begin{align}
\label{eq:downlink_channels}
	\widehat{\mathbf{H}}_{:,T_u+n}=g_{\bm{\phi}_2}(\mathbf{z}_n)\ \text{for}\ n=0,1,\cdots,T_d-1.
\end{align}

The structures of all the FNNs are listed in TABLE \ref{table_FNN_structure}.

\begin{table}[!t]
	\centering
	\renewcommand{\arraystretch}{1}
	\caption{Structures of the FNNs.}
	\label{table_FNN_structure}
	\begin{tabular}{c|c|c|c}
		\hline
		&\makecell[c]{Layer} &\makecell[c]{Output size} &\makecell[c]{Activation}\\
        \hline
		\multirow{5}*{$f_{\bm{\theta}_1}(\cdot)$ \& $f_{\bm{\theta}_2}(\cdot)$}
		&FC &40 &Tanh\\
		\cline{2-4}
		&FC &40 &Tanh\\
		\cline{2-4}
		&FC &40 &Tanh\\
		\cline{2-4}
		&FC &40 &Tanh\\	
		\cline{2-4}
		&FC &40 &-\\	
		\hline
		\multirow{2}*{$g_{\bm{\phi}_1}(\cdot)$}
		&FC &40 &Tanh\\
		\cline{2-4}
		&FC &$2L$ &-\\	
		\hline
		\multirow{1}*{$g_{\bm{\phi}_2}(\cdot)$}
		&FC &$2N$ &-\\
		\hline
		\multirow{2}*{$f_u(\cdot)$ \& $f_r(\cdot)$}
		&FC &40 &Tanh\\
		\cline{2-4}
		&FC &$2L$ &Sigmoid\\	
		\hline
		\multirow{2}*{$f_i(\cdot)$}
		&FC &40 &Tanh\\
		\cline{2-4}
		&FC &$2L$ &-\\	
		\hline
	\end{tabular}	
\end{table}

\subsection{Training Procedure}
\label{Training}

Denote $\mathcal{D}$ as the network training dataset, where $|\mathcal{D}|=N_{tr}$ is the number of training samples.
One sample in $\mathcal{D}$ is an input-label pair $(\mathbf{H}_{\mathcal{M},\mathcal{T}_u},\mathbf{H}_{:,\mathcal{T}_d})$.
More details about the dataset generation are presented in Section \ref{Simulation}.
The set of parameters for all the NNs is denoted as $\mathcal{P}=\{\bm{\theta}_1,\bm{\theta}_2,\bm{\phi}_1,\bm{\phi}_2,\bm{\psi}\}$.
To train the LODE-CENet, we adopt the end-to-end learning for all the parameters in $\mathcal{P}$.
The loss function is defined as the mean square error (MSE) between the output of LODE-CENet and the label, which is written as
\begin{align}
	\mathcal{L}=\frac{1}{M_{tr}NT_d}\sum_{m=1}^{M_{tr}}\left\|[\mathbf{H}_{:,\mathcal{T}_d}]_m-[\widehat{\mathbf{H}}_{:,\mathcal{T}_d}]_m\right\|_F^2,
\end{align}
where $M_{tr}$ is the batch size for training and $[\cdot]_m$ represents the $m$-th sample in the mini-batch.
The \emph{AdaMax} optimizer, a variant of the adaptive moment optimizer \cite{adam}, is adopted to optimize the parameters in $\mathcal{P}$.

\section{Simulation Results}
\label{Simulation}

\begin{table}[!t]
	\centering
	\renewcommand{\arraystretch}{1.2}
	\caption{Simulation Parameters.}
	\label{table_simulation_parameters}
	\begin{tabular}{c|c}
		\hline
		Number of BS antennas $N$ &64 \\
        \hline
		Carrier frequency $f_c$ &60 GHz \\
		\hline
		BS antenna spacing $d$ &$\lambda/2$ \\
		\hline
		Channel scattering paths $N_p$ &6 \\
        \hline
		User velocity $v$ &70 km/h \\
        \hline
		Channel coherence interval $L_c$ &50 \\
		\hline
		Sampling period $T_s$ &0.05 us\\
		\hline
	\end{tabular}	
\end{table}

\begin{figure*}
	\begin{minipage}[t]{60mm}
	\centering
	\includegraphics[width=60mm]{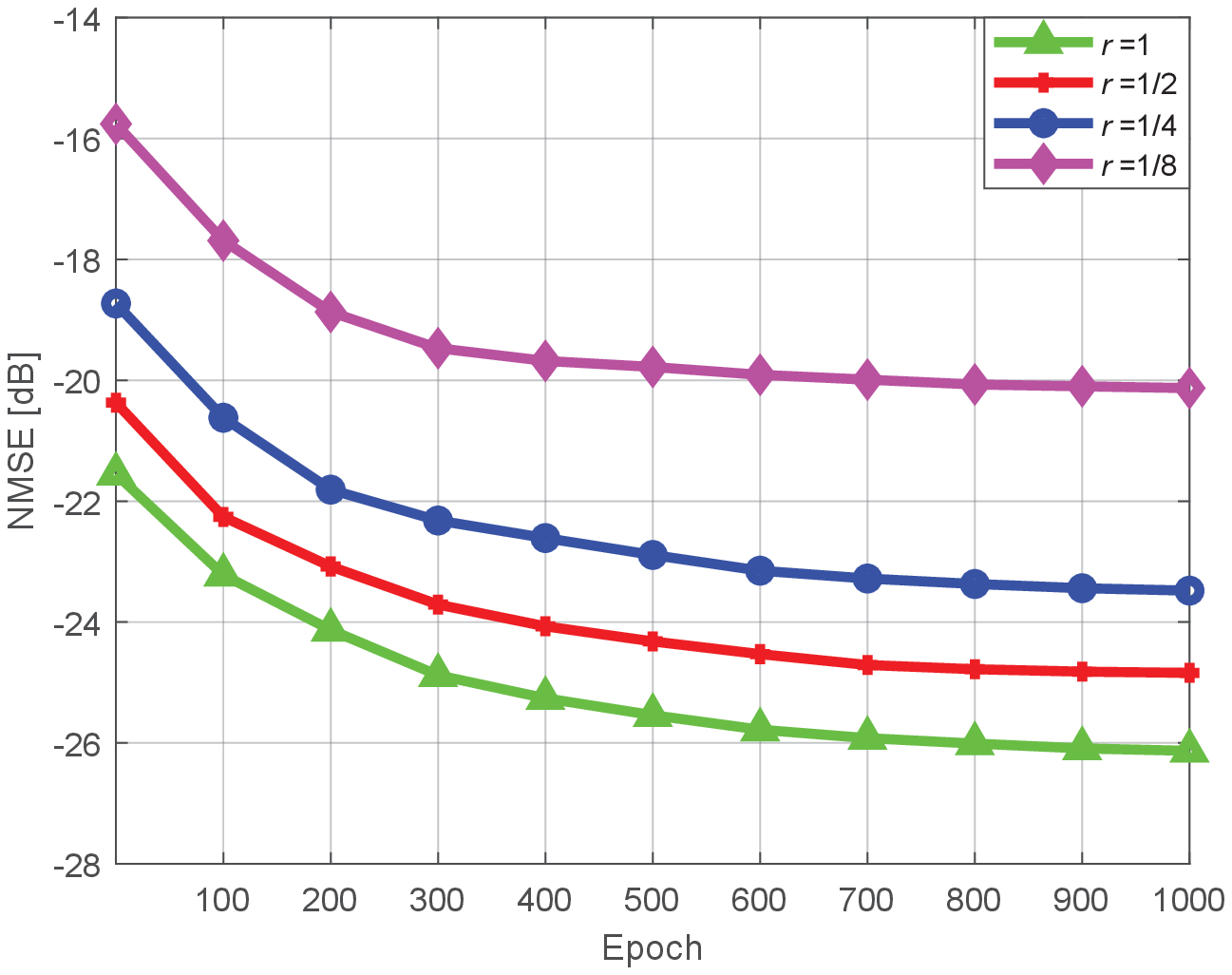}
	\caption{The NMSE of channel extrapolation versus epoch for different $r$ values.}
	\label{Epoch_NMSE_SNR}
	\end{minipage}
	\begin{minipage}[t]{60mm}
	\centering
	\includegraphics[width=60mm]{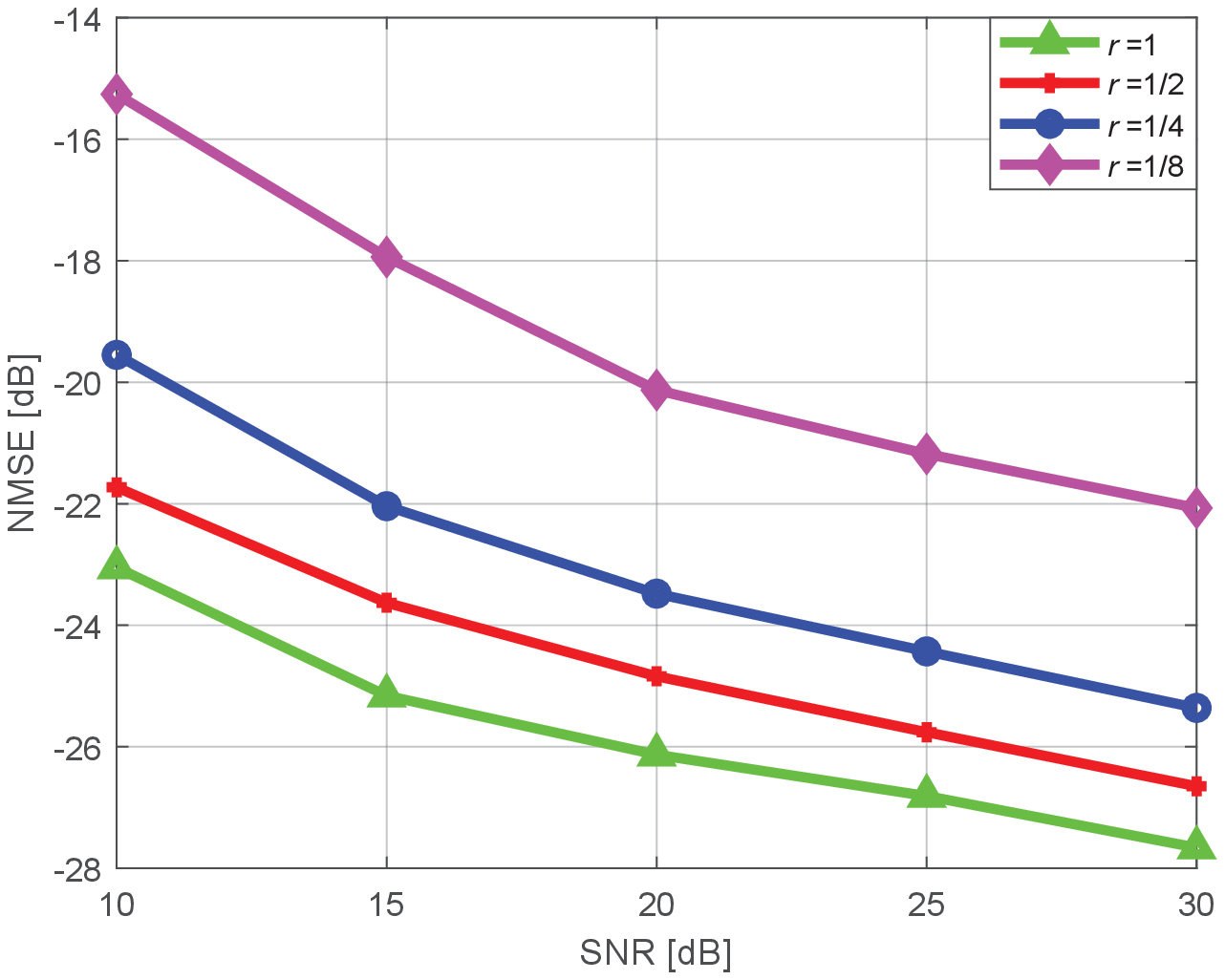}
	\caption{The NMSE of channel extrapolation versus SNR for different $r$ values.}
	\label{SNR_NMSE_r}
	\end{minipage}
	\begin{minipage}[t]{60mm}
	\centering
	\includegraphics[width=60mm]{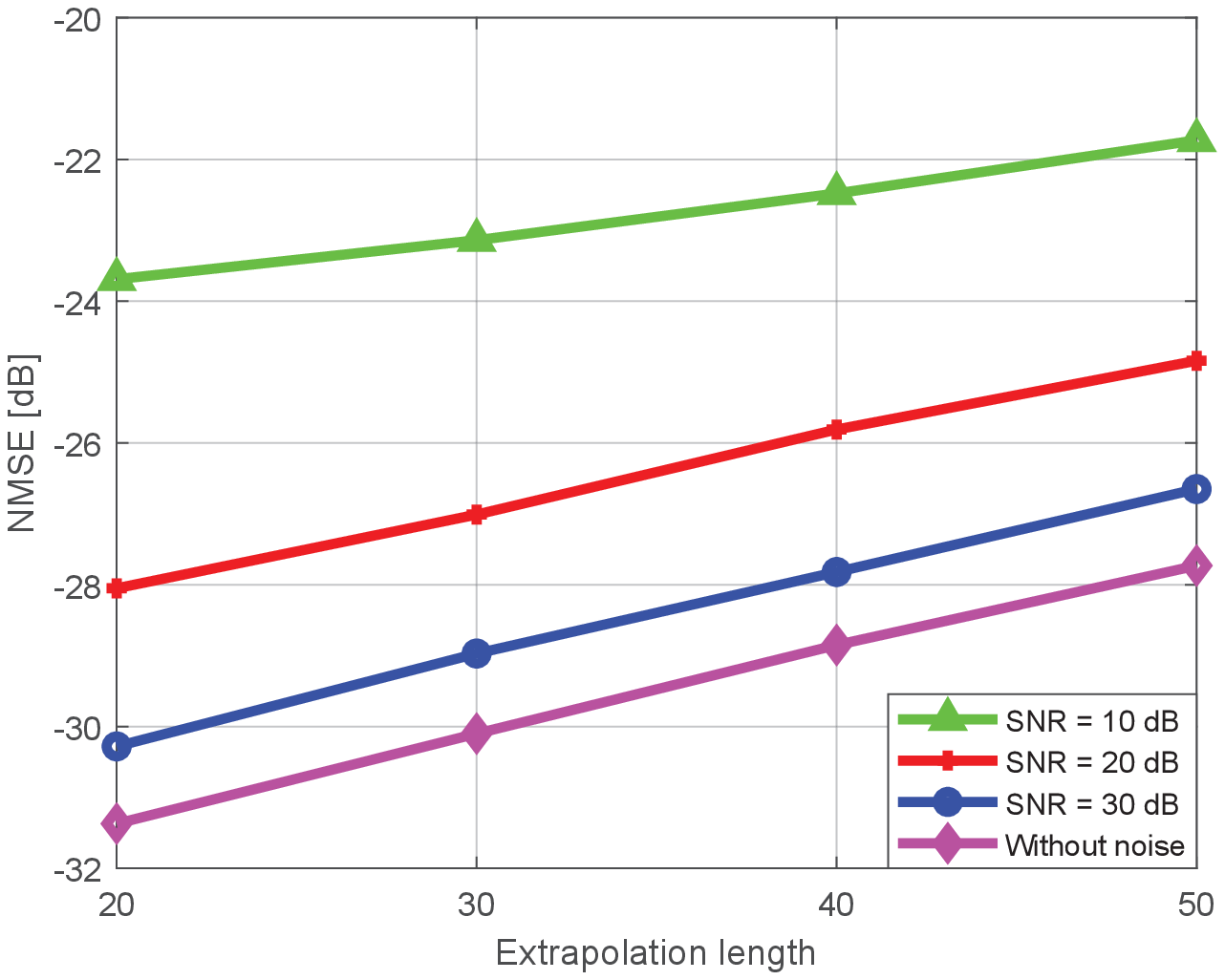}
	\caption{The NMSE of channel extrapolation versus time domain extrapolation length for different SNRs.}
	\label{Len_NMSE_SNR}
	\end{minipage}
\end{figure*}

To evaluate the performance of the designed network, we resort to the DeepMIMO dataset \cite{DeepMIMO}, which is widely used in DL applications for massive MIMO systems.
The outdoor ray-tracing scenario `O1' of the DeepMIMO dataset is adopted and the BS 1 in the `O1' scenario is set as the ULA with 64 antennas.
We summarize the simulation parameters in TABLE \ref{table_simulation_parameters}.
To generate the channel dataset, we first fix the user velocity $v$.
The Doppler shift is $\nu_p=\frac{v}{\lambda}\cos\vartheta_p$, where $\vartheta_p$ denotes the angle between the user movement direction and the $p$-th path.
To reduce the size of the dataset and save the memory, we select $\vartheta_p\in[-20^{\circ},20^{\circ}]$.
Then, the parameters $\alpha_p$, $\nu_p$, $\tau_p$ and $\theta_p$ of each user can be obtained form the DeepMIMO dataset in the `O1' scenario.
With the parameters $\{\alpha_p, \nu_p, \tau_p, \theta_p\}_{p=1}^{N_p}$, $L_c$, $f_c$ and $T_s$, we can generate the time-varying channel sample of each user according to \eqref{eq:channel_model} for $n\in\mathcal{T}$.
We select the users located in the region from the 401-th row to the 510-th row.
Since each row in the user's region contains 181 users in the `O1' scenario, the total number of the channel sample is 19910.
We employ 80\% of the dataset for training and the rest for validation.
Specifically, the dimension of the hidden state is the same as that of the latent state and is set as $L=48$.
The total epochs for training is 1000.
We set the initial learning rate as 0.004 and halve it every 50 epochs.
The AdaMax optimizer is adopted for network training with batch size $M_{tr}=80$.
To evaluate the channel extrapolation performance, we use the normalized MSE (NMSE) $\mathbb{E}\left\{\frac{\|\mathbf{H}_{:,\mathcal{T}_d}-\widehat{\mathbf{H}}_{:,\mathcal{T}_d}\|_F^2}{\|\mathbf{H}_{:,\mathcal{T}_d}\|_F^2}\right\}$ as the metric.

Fig. \ref{Epoch_NMSE_SNR} shows the channel extrapolation performance of the designed LODE-CENet on the validation set versus the training epochs.
We set $T_u=T_d=50$ and SNR=20 dB.
Four different spatial compression ratio, i.e., $r=1, \frac{1}{2}, \frac{1}{4}, \frac{1}{8}$, are considered for comparison.
It can be seen that the NMSE decreases with the training epochs, and the extrapolation performance of the designed network almost achieve the convergence after 700 epochs.


Fig. \ref{SNR_NMSE_r} depicts the extrapolation performance of the designed LODE-CENet on the validation set versus the SNR.
We set $T_u=T_d=50$ and investigate the network performance under four different $r$, i.e., $r=1$, $\frac{1}{2}$, $\frac{1}{4}$, $\frac{1}{8}$.
It can be seen that the NMSE decreases as the SNR increases, which indicates that the network input data closer to the real channel information is beneficial to improve the extrapolation performance of the designed network. 
Furthermore, as we can see from both Fig. \ref{Epoch_NMSE_SNR} and Fig. \ref{SNR_NMSE_r}, the gap between the curves with $r=\frac{1}{8}$ and the one with $r=\frac{1}{4}$ is significantly larger than that of the other adjacent curves; in other words, as the spatial compression ratio drops below $\frac{1}{4}$, it becomes more and more difficult for the designed LODE-CENet to capture the spatial correlation between the channels at different antennas.


Fig. \ref{Len_NMSE_SNR} shows the network extrapolation performance on the validation set versus the time domain extrapolation length, i.e., $T_d$.
Specifically, $T_u$ and $r$ are set to $50$ and $\frac{1}{2}$, respectively, and three different SNRs, i.e., 10 dB, 20 dB, 30 dB and no noise are considered.
As can be seen in Fig. \ref{Len_NMSE_SNR}, it is intuitive that the channel extrapolation performance becomes better as the SNR increases.
Moreover, the NMSE increases as the extrapolation length increases, which indicates that it becomes more and more difficult to capture the evolution of the channels in time domain with the increase of time.


\section{Conclusion}
\label{Conclusion}

In this paper, we have considered the DL-based downlink channel acquisition in a time-varying massive MIMO system. 
Specifically, the hybrid precoding structure is adopted at the BS to reduce the number of active antennas.
Furthermore, a latent ODE-based network has been designed under the VAE framework to perform the antenna-time domain channel extrapolation from the partial uplink channels to the full downlink channels, where the GRU is adopted for the encoder and the FNN is used for the decoder.
Simulation results have shown that the designed network can efficiently infer the full downlink channels, which reveals that the proposed scheme can significantly reduce the channel training overhead.

\vspace{-1mm}
\linespread{1}
\balance

\end{document}